\documentclass[%
 reprint,
 amsmath,amssymb,
 aps,
]{revtex4-2}

\usepackage{graphicx}
\usepackage{dcolumn}
\usepackage{bm}

\begin{document}

\preprint{APS/123-QED}

\title{A Paradigm for Density Functional Theory Using \\Electron Distribution on the Energy Coordinate}

\author{Hideaki Takahashi}
 \email{ hideaki.takahashi.c4@tohoku.ac.jp}
\affiliation{%
Department of Chemistry, Graduate School of Science,\\
 Tohoku University, Sendai, Miyagi 980-8578, Japan 
}%




\date{\today}

\begin{abstract}
Static correlation error(SCE) inevitably emerges when a dissociation of a covalent bond is described with a conventional denstiy-functional theory (DFT) for electrons. SCE gives rise to a serious overshoot in the potential energy at the dissociation limit even in the simplest molecules. The error is attributed to the basic framework of the approximate functional for the exchange correlation energy $E_{xc}$ which refers only to local properties at coordinate $\bm{r}$, namely, the electron density $n(\bm{r})$ and its derivatives. To solve the problem we developed a functional $E^e_{xc}$ which uses the energy electron distribution $n^e(\epsilon)$ as a fundamental variable in DFT. $n^e(\epsilon)$ is obtained by the projection of the density $n({\bm{r}})$ onto an energy coordinate $\epsilon$ defined with the external potential of interest. 
The functional was applied to the dissociations of single, double, and triple bonds in small molecules showing reasonable agreements with the results given by a high level molecular orbitals theory. We also applied the functional to the computation of the energy change associated with spin depolarization and symmetrization in Carbon atom, which made an improvement over the conventional functional.  This work opens the way for development of tougher functional that necessitates non-local properties of electrons such as kinetic energy functional.   　  
\end{abstract}

\maketitle


\section{\label{sec:level1}The static correlation error in density functional theory}

The density-functional theory (DFT) offers a versatile framework to study properties 
of interacting many-particles systems such as liquids or electrons in molecules or bulk 
materials\cite{rf:Hansen2013, rf:parr_yang_eng}. DFT owes its success in physics to the 
simple treatments of the complex correlations among the particles in terms of their distribution functions. 
For instance free energy change associated with an insertion of a solute into a pure solvent can be formulated by a
functional of the distribution $\rho$ of the solvent around the solute. In the electronic structure calculations
the electron density $n$ plays an essential role in describing the exchange and correlation energies $E_{xc}$ of electrons. 
The key approach common to these theories is to project the effect of the particles correlations onto the
1-body potential given as a functional of the distribution $\rho$ or of the electron density $n$. The construction of the functionals is, of course, based on the statistical mechanics in the theory of solutions and on the quantum mechanics in the DFT for electrons. \\
\indent The DFT for electrons, Kohn-Sham(KS) DFT\cite{rf:kohn1965pr} in particular, has been extensively utilized in the field of physics and chemistry.  A vast amount of work has been devoted to the foundation of the theoretical basis of the DFT and the developments of the functionals as well as their applications. Actually, KS-DFT has been established as a reliable and efficient method in the electronic structure calculation.  However, the functionals for KS-DFT still suffer from two major problems\cite{Cohen2008}. First, in the approximate DFT functional, the self-interaction of an electron due to Coulomb interaction does not fully cancel the exchange energy of itself, that is often referred to as self-interaction error (SIE). Second, static correlation error (SCE) gives rise to erroneous destabilization in energy in describing the dissociation of a chemical bond when spin-symmetry adapted wave functions are used. Indeed, these errors seriously affect the energetics even in the simplest molecules. In the dissociation of the H$_2$ molecule for instance, a GGA(generalized gradient approximation)\cite{rf:parr_yang_eng} level $E_{xc}$ functional overshoots the potential energy at the dissociation by $\sim 50$ kcal/mol\cite{rf:becke2003jcp}. The SCE arises more or less in stretching every covalent bond. The major objective of the present Letter is to develop a simple and efficacious functional to solve the SCE problem on the basis of a new DFT paradigm where the electron distribution on the \textit{energy coordinate} plays a role.  \\
\indent The mechanism underlying the emergence of SCE can be elucidated in terms of the exchange hole model used in the local density approximation (LDA)\cite{rf:parr_yang_eng} which constitutes the basis of the developments of all the $E_{xc}$ functionals. The exchange hole in the homogeneous electron gas (HEG) is in particular the most prevailing hole model.  Here, we consider the bond dissociation of an H$_2$ molecule for simplicity. When the bond is infinitely stretched with the spin symmetry maintained, the population of the electron with a spin $\sigma$ on each atom becomes half an electron. Hence, the hole depth modeled by the LDA approach is also just half of the exact exchange hole since the depth of the hole is equal to the density $n_\sigma(\bm{r})$, which causes serious underestimation in the exchange energy. One way to compensate the error is to incorporate some fraction of the exchange hole of the opposite spin into the total hole\cite{rf:becke2003jcp} although there is no theoretical justification for the treatment.  
\begin{figure}[h]
\centering
\scalebox{0.35}[0.35] {\includegraphics[trim=70 80 80 90,clip]{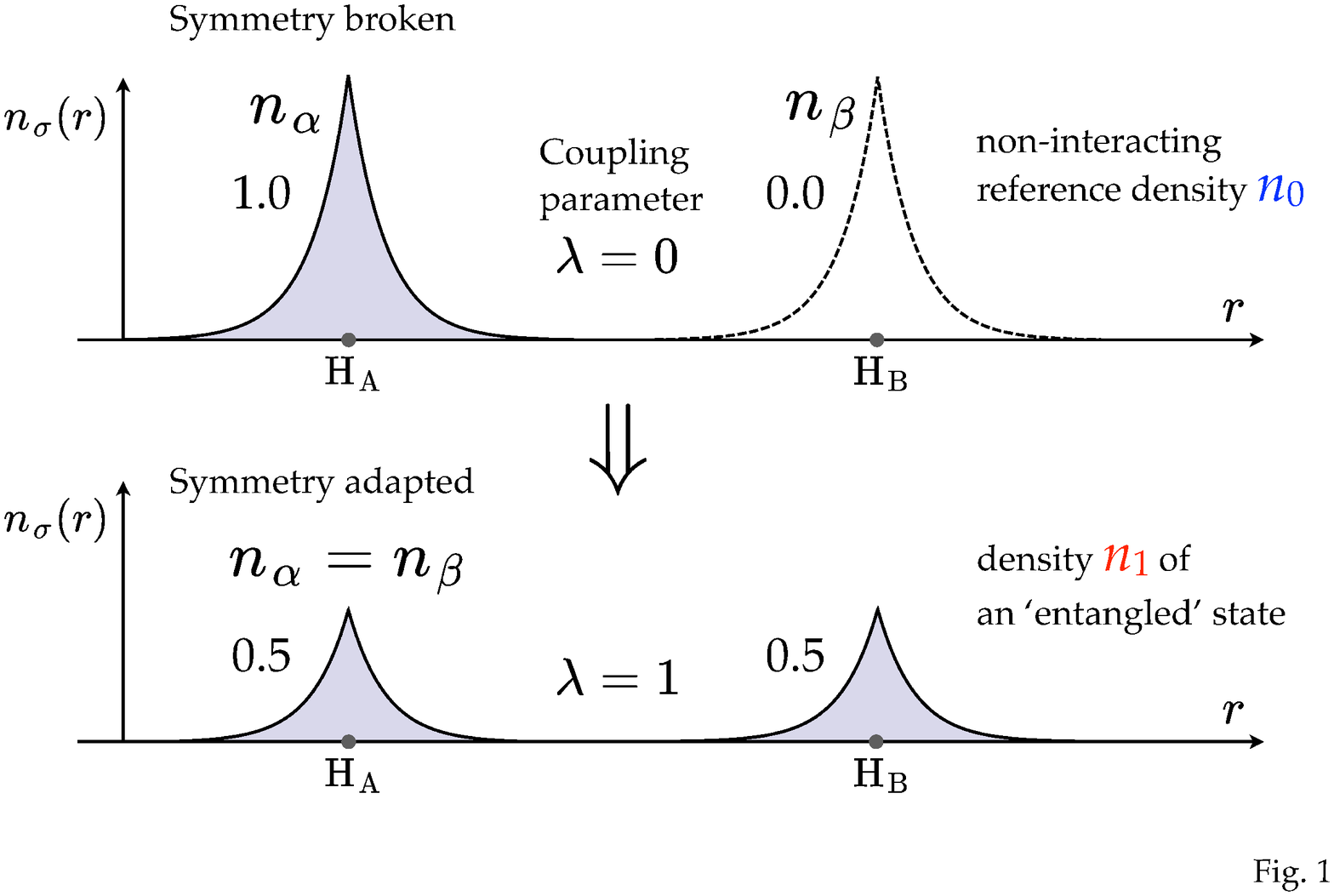}}            
\caption{\label{Coupling}Electron densities $n_\sigma (\sigma = \alpha, \beta)$ of an H$_2$ molecule at the dissociation limit. The overlap of the electron densities of independent H atoms (upper) and the density of the entangled H atoms (lower). The shaded regions show the electron distributions for spin $\alpha$. The real numbers are fractional populations with $\alpha$ spin. The coupling parameter $\lambda$ represents the gradual variation of the density from $n_0$ to $n_1$. }
\end{figure} 
\\
\section{\label{sec:level2}Formulation of a functional with a distribution on the energy coordinate }

\indent To clarify the properties of the functional to be developed in this Letter we present in Fig. \ref{Coupling} the gradual variation of the electron density with spin $\sigma$ from a reference density $n_0$ to $n_1$ of the interacting system through a coupling parameter $\lambda (0 \le \lambda \le 1)$ for a hydrogen molecule when the bond distance $R$ is infinitely extended($R=\infty$). $n_0$ is merely the overlap of the electron densities with opposite spins for the independent H atoms placed at sites H$_\text{A}$ and H$_\text{B}$. $n_1$ is the spin restricted density of the \textquoteleft entangled\textquoteright\; H atoms. In the following we drop the spin index for the sake of notational simplicity unless otherwise stated. We also assume that the reference density $n_0$ is given from the outset. It is apparent that the spatial behavior of the electron density $n_1(\bm{r})$ completely differs from that of $n_0(\bm{r})$. The target exchange-correlation energy $\overline{E}_{xc}[n_1]$ can be expressed in the form of coupling parameter integration, thus,      
\begin{equation}
\begin{aligned} 
\overline{E}_{x c}\left[n_{1}\right] &=E_{x c}\left[n_{0}\right]+\int_{0}^{1} d \lambda \frac{d \widetilde{E}_{x c}\left[n_{\lambda}\right]}{d \lambda} \\ 
&=E_{x c}\left[n_{0}\right]+\int_{0}^{1} d \lambda \int d \bm{r} \frac{d n_{\lambda}(\bm{r})}{d \lambda} \frac{d \widetilde{E}_{x c}\left[n_{\lambda}\right]}{d n_{\lambda}(\bm{r})} 
\end{aligned}
\label{eq:coupling_r}
\end{equation}
where $E_{xc}$ is a certain conventional functional and $\widetilde{E}_{x c}$ represents a functional specific to the evaluation of the difference in the exchange-correlation energy from $\lambda = 0$ to $1$. Of course, the integration in Eq. (\ref{eq:coupling_r}) should be evaluated as zero when the exact exchange-correlation functional $E_{xc}^\text{ex}$ is employed($\widetilde{E}_{xc}=E_{xc}^\text{ex}$). It is also important to note that any intermediate density expressed as $n_\lambda = (1-\lambda) n_0 + \lambda n_1$ should also give the same result, i.e. $\overline{E}_{xc}[n_\lambda]=\overline{E}_{xc}[n_0]=\overline{E}_{xc}[n_1]\; (0 < \lambda < 1)$. This imposes a quite tough requirement on the functional $\widetilde{E}_{x c}$. 
\\
\indent Here, we introduce the energy electron density $n^e(\epsilon)$\cite{rf:Takahashi2018} by projecting the density $n(\bm{r})$
\begin{equation}
n^e(\epsilon) = \int d\epsilon\; n(\bm{r})\delta(\epsilon - \upsilon(\bm{r}))
\label{eq:EDF}
\end{equation}
where $\upsilon(\bm{r})$ is the external potential of interest used to define the energy coordinate $\epsilon$. Explicitly, $\upsilon(\bm{r})$ is given by $\upsilon(\bm{r}) = \sum_A Z_A / \left|\bm{r} - \bm{R}_A \right| $, where $Z_A$ and $\bm{R}_A$ are charge and position of nuclear A in the molecule. As was proved in the previous work\cite{rf:Takahashi2018}, there exists one-to-one correspondence between a subset of the external potentials and a subset of the energy electron distributions. Further, a rigorous framework of the density functional theory can be established almost in parallel to the conventional DFT for electrons\cite{rf:Takahashi2018}.    
\begin{figure}[h]
\centering
\scalebox{0.36}[0.36] {\includegraphics[trim=70 120 80 90,clip]{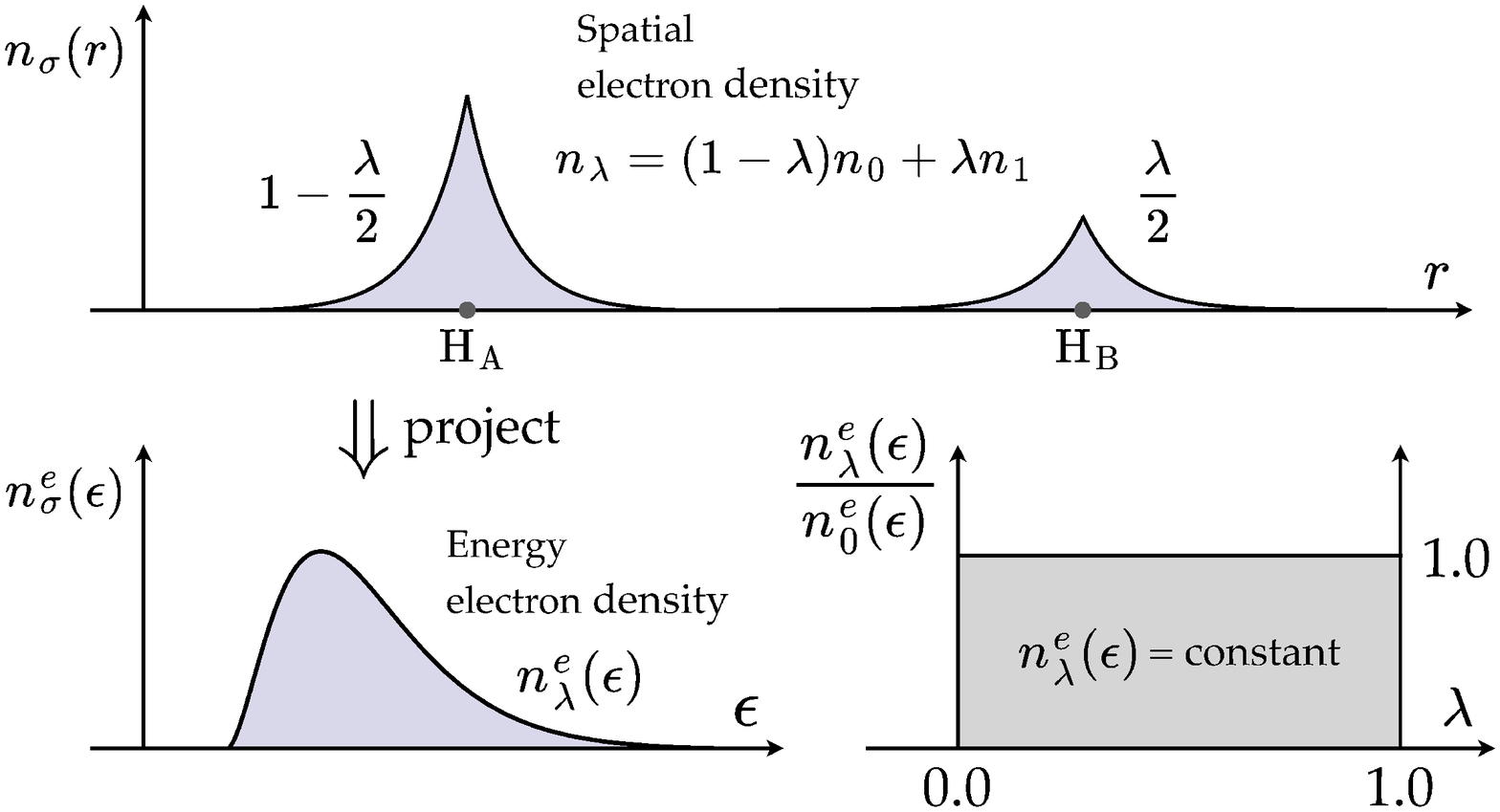}}            
\caption{\label{Projection} The spatial electron density $n_\lambda(\bm{r})$ of an H$_2$ molecule(upper) and its projection $n_\lambda^e(\epsilon)$ onto the energy coordinate $\epsilon$(lower left). At the dissociation limit $n_\lambda^e(\epsilon)$ stays constant with respect to the variation of $\lambda$ (lower right). }
\end{figure} 
The advantage of using the distribution $n^e(\epsilon)$ is that $n_{\lambda}^e(\epsilon)$ stays constant for an arbitrary $\epsilon$ with respect to the variation of $\lambda$ at $R=\infty$. Hereafter, we attach the superscript $e$ on the functions and functionals that are relevant to the energy coordinate to avoid confusions.   
We replace the distribution $n_{\lambda}(\bm{r})$ in the integrand by the corresponding energy electron density $n_{\lambda}^e(\epsilon)$, which reads 
\begin{equation}
\overline{E}_{x c}\left[n_{1}\right] =E_{x c}\left[n_{0}\right]+\int_{0}^{1} d \lambda \int d \epsilon \frac{d n_{\lambda}^e(\epsilon)}{d \lambda} \frac{d \widetilde{E}_{x c}^e\left[n_{\lambda}^e\right]}{d n_{\lambda}^e(\epsilon)} 
\label{eq:coupling_e}
\end{equation}
Assuming the linear dependence of $n_{\lambda}^e(\epsilon)$ on $\lambda$, we have 
\begin{equation}
\begin{aligned} 
\overline{E}_{x c}\left[n_{1}\right]=& E_{x c}\left[n_{0}\right] \\ &+\int d \epsilon\left(n_{1}^e(\epsilon)-n_{0}^e(\epsilon)\right) \int_{0}^{1} d \lambda\; \widetilde{\upsilon}_{x c}^{e}\left[n_{\lambda}](\epsilon)\right.
\end{aligned}
\label{eq:coupling_e2}
\end{equation}
where $\widetilde{\upsilon}_{x c}^{e}\left[n_{\lambda}^e](\epsilon)\right.$ is the exchange-correlation potential defined on the energy coordinate. Obviously, the integral in Eq. (\ref{eq:coupling_e2}) becomes zero at the dissociation limit for any choice of the approximate functional $\widetilde{E}_{x c}^e\left[n^e\right]$ since the relation $n_1^e(\epsilon)-n_0^e(\epsilon) = 0$ is fully satisfied. Even in a weakly interacting system it is anticipated that $n_1^e$ is very close to $n_0^e$ and, hence, the integral is evaluated to be a finite but small value. The energy electron distribution $n^e(\epsilon)$ also provides a benefit in the evaluation of the coupling parameter integration for $\widetilde{\upsilon}_{x c}^{e}\left[n_{\lambda}^e]\right.$. The weak dependence of $n_{\lambda}^e$ on $\lambda$ justifies the application of the linear response scheme in constructing $\widetilde{\upsilon}_{xc}^e[n_1^e](\epsilon)$ from $n_0^e$, 
\begin{equation}
\begin{aligned}
\overline{\upsilon}_{x c}\left[n_{1}^{e}\right](\epsilon)=& \upsilon_{x c}\left[n_{0}^{e}\right](\epsilon) \\ &+\int d \epsilon^{\prime} \frac{\delta \widetilde{\upsilon}_{x c}^{e}\left[n_{0}^{e}\right](\epsilon)}{\delta n_0^{e}\left(\epsilon^{\prime}\right)}\left(n_{1}^{e}\left(\epsilon^{\prime}\right)-n_{0}^{e}\left(\epsilon^{\prime}\right)\right) 
\end{aligned}
\label{eq:LR}
\end{equation}
Thus, the exchange-correlation potential $\overline{\upsilon}_{x c}\left[n_{1}^{e}\right]$ can be constructed only from the knowledge of the reference density $n_0^e$ and the response function at $n_0^e$. Equation (\ref{eq:LR}) allows us to determine $\overline{\upsilon}_{x c}\left[n_{1}^{e}\right]$ as well as $n_1^e$ in a self-consistent manner\cite{rf:Takahashi2018}. \\
\indent Now, we formulate a functional $\overline{E}_{xc}[n_1]$ provided that the exact spin-restricted density $n_1$ is given. Without the coupling parameter, $\overline{E}_{xc}[n_1]$ in Eq. (\ref{eq:coupling_e}) can be merely expressed as
\begin{equation}
\overline{E}_{x c}\left[n_{1}\right]=E_{x c}\left[n_{0}\right]+\widetilde{E}_{x c}^{e}\left[n_{1}^{e}\right]-\widetilde{E}_{x c}^{e}\left[n_{0}^{e}\right]
\label{eq:Exc_static}
\end{equation} 
We define in Eq. (\ref{eq:Exc_static}) the static correlation $E_\text{sc}$ as $E_\text{sc} = E_{x c}\left[n_{0}\right]-\widetilde{E}_{x c}^{e}\left[n_{0}^{e}\right]$ since $\widetilde{E}_{x c}^{e}\left[n_{1}^{e}\right]$ itself describes the $E_{xc}$ energy of the system with the density $n_{1}^{e}$. As discussed above Eq. (\ref{eq:Exc_static}) realizes $\overline{E}_{x c}\left[n_{1}\right] = E_{x c}\left[n_{0}\right]$ when $R = \infty$ since $n_{1}^{e} = n_{0}^{e}$ is ensured.  In other words, Eq. (\ref{eq:Exc_static}) works exactly when the corresponding exchange hole is equally fragmented onto the two atomic sites. Importantly, at the other limit where the two atomic sites coincide($R=0$) $E_\text{sc}=0$ is attained under an appropriate choice of $\widetilde{E}_{x c}^{e}$. However, in a system with an intermediately dissociated bond full inclusion of $E_\text{sc}$ gives rise to an artificial lowering of $\overline{E}_{xc}$ as demonstrated in Ref. \onlinecite{rf:Takahashi2018}.   \\
\indent A reasonable way to solve the problem is to attenuate the energy $E_\text{sc}$ in the intermediate region according to the hole delocalization. To this end we first rewrite Eq. (\ref{eq:Exc_static}) equivalently in terms of the exchange-correlation energy density $U_{xc}$, 
\begin{equation}
\begin{aligned} \overline{E}_{x c}\left[n_{1}\right] & = \int d \epsilon\; U_{x c}^{e}\left[n_{1}^{e}\right](\epsilon) n_{1}^{e}(\epsilon) \\
+ \int d \epsilon &\left(\widetilde{U}_{x c}\left[n_{0}\right](\epsilon)-\widetilde{U}_{x c}^{e}\left[n_0^{e}\right](\epsilon)\right)n_{0}^{e}(\epsilon) \\ 
\end{aligned}
\label{eq:Exc_static2}
\end{equation}
where $U^e_{xc}(\epsilon)$ is the exchange-correlation energy density at $\epsilon$ and $\widetilde{U}_{x c}\left[n_{0}\right](\epsilon)$ is specifically the weighted average of $U_{xc}[n_0](\bm{r})$ over the hypersurface with the energy coordinate $\epsilon$ and defined as 
\begin{equation}
\widetilde{U}_{xc}\left[n_{0}\right](\epsilon)=\frac{1}{n_{0}^{e}(\epsilon)}\int U_{xc}\left[n_{0}\right](\bm{r}) n_{0}(\bm{r}) \delta\left(\epsilon-\upsilon(\bm{r})\right) d \bm{r}
\end{equation}
It is easy to see that $\int d \epsilon\; \widetilde{U}_{x c}\left[n_{0}\right](\epsilon) n_{0}^{e}(\epsilon)=E_{x c}\left[n_{0}\right]$.
Then, we introduce a factor $\mathcal{D}^e(\epsilon)$ as a function of the energy coordinate $\epsilon$. $\mathcal{D}^e(\epsilon)$ measures the degree of the fragmentation of the exchange hole over multiple sites. Using the factor Eq. (\ref{eq:Exc_static2}) is to be revised as 
\begin{equation}
\begin{aligned} \overline{E}_{x c}\left[n_{1}\right] & = \int d \epsilon\; U_{x c}^{e}\left[n_{1}^{e}\right](\epsilon) n_{1}^{e}(\epsilon) \\
+ \int d \epsilon &F\left(\mathcal{D}^e(\epsilon)\right)\left(\widetilde{U}_{x c}\left[n_{0}\right](\epsilon)-\widetilde{U}_{x c}^{e}\left[n_0^{e}\right](\epsilon)\right)n_{0}^{e}(\epsilon) \\ 
\end{aligned}
\label{eq:Exc_static_rev}
\end{equation}
where $F$ is a function of $\mathcal{D}^e(\epsilon)$.  Equation (\ref{eq:Exc_static_rev}) shows that the static correlation $E_\text{sc}$ is incorporated into $\overline{E}_{xc}$ according to the weight $F(\mathcal{D}^e(\epsilon))$. $F$ is supposed to be a monotonically increasing function ranging from 0 to 1 and can be designed to realize $E_\text{sc}$ appropriately in the intermediate region.  
\section{\label{sec:level3}Numerical implementation of the functional }
\begin{figure}[h]
\centering
\scalebox{0.49}[0.49] {\includegraphics[trim=160 120 80 130,clip]{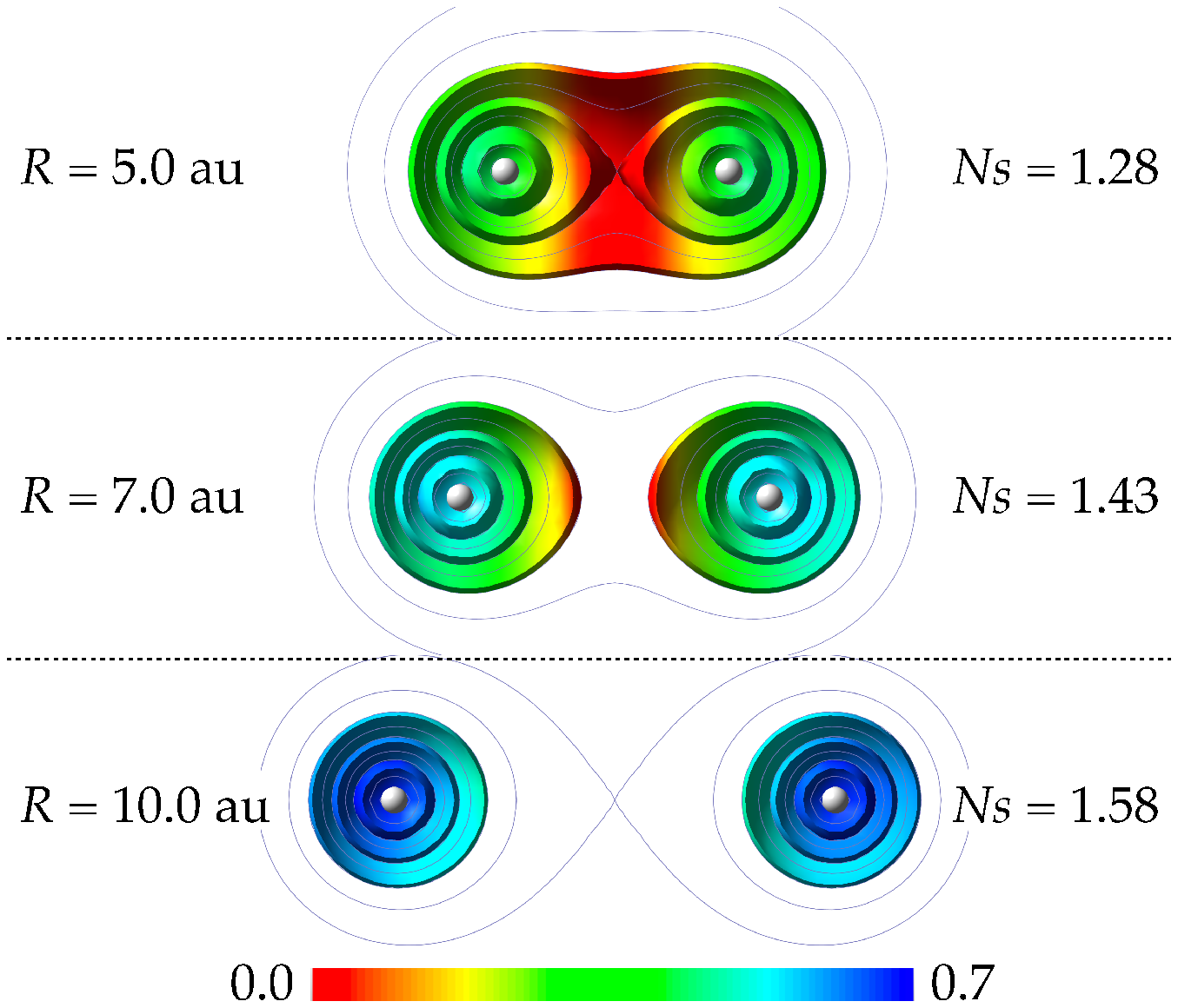}}            
\caption{\label{Deloc} The delocalization factor $\mathcal{D}(\bm{r})$ mapped onto the isosurfaces of the external potential $\epsilon=\upsilon(\bm{r})$ for H$_2$ molecule with various bond distances $R$. Depicted isosurfaces are for $\epsilon = 0.6, 0.8$, and $1.2$ au. The contours of $\epsilon$ on the molecular plane are also drawn with the gray thin lines ($\epsilon$ = 0.4, 0.5, 0.7, 1.0, 1.2, 2.0, and 4.0 au). }
\end{figure}

We implement a numerical functional $\overline{E}_{xc}$ with the form of Eq. (\ref{eq:Exc_static_rev}) on the basis of the conventional DFT functional. To this end, we adapt BLYP\cite{rf:becke1988pra, rf:lee1988prb} known as a representative GGA functional to evaluate $\widetilde{E}_{xc}^e$ in terms of the energy electron density $n^e(\epsilon)$. The variables $G[n(\bm{r})]$ needed as arguments for the BLYP functional, namely, $n(\bm{r})$ itself and its derivatives ($|\nabla n(\bm{r})|$ and $\nabla^2 n(\bm{r})$) are transformed to the corresponding quantities $G^e[n^e(\epsilon)]$ defined on the energy coordinate. Explicitly, $G^e[n^e(\epsilon)]$ are obtained as the average of $G[n(\bm{r})]$ over the hypersurface with energy coordinate $\epsilon$, thus, 
\begin{equation}
G^e[n^e(\epsilon)] = \Omega(\epsilon)^{-1}\int d\bm{r} G[n(\bm{r})]\delta(\epsilon - \upsilon(\bm{r}))
\label{eq:GGA_e}
\end{equation} 
where $\Omega(\epsilon)$ is the volume of the region with coordinate $\epsilon$ and defined by
\begin{equation}
\Omega(\epsilon) = \int d\bm{r}\; \delta(\epsilon - \upsilon(\bm{r}))
\label{eq:Omega}
\end{equation} 
It is readily recognized that each quantity $G^e[n^e(\epsilon)]$ has the same dimension with the original variable $G[n(\bm{r})]$. Therefore, it is possible to substitute $G^e[n^e(\epsilon)]$ for $G[n(\bm{r})]$ in adopting the conventional GGA.  Of course, this is not a unique way of developing the functional for $n^e$. However, it will be an instant but reasonable approach to realize the functional based on the energy coordinate. It should be stressed that $\widetilde{E}_{xc}^e$ with the variables on the energy coordinate (Eqs. (\ref{eq:GGA_e}) and (\ref{eq:Omega})) provides exactly the same results with the original GGA functional in principle when it is applied to hydrogenic atoms, i.e. spherically symmetric 1-electron systems, because $G[n(\bm{r})]$ are constant over a contour surface with an energy coordinate.   \\
\indent Evaluation of the delocalization factor $\mathcal{D}^e(\epsilon)$ and the determination of the correlation factor $F$ in Eq. (\ref{eq:Exc_static_rev}) are the keys to the success of the application of the present theory. In this work we utilize the generalized Becke-Roussel(GBR)\cite{rf:becke2003jcp} exchange hole for the evaluation of $\mathcal{D}^e(\epsilon)$. In the GBR approach the exchange hole at a reference point $\bm{r}$ is modeled by the hole for a hydrogenic atom. Explicitly, the hole function $n_x^\text{GBR}(A,\zeta, r)$ is expressed by a Slater function,
\begin{equation}
n_x^\text{GBR}(A,\zeta, r) = A\exp(-\zeta r)
\label{eq:GBR}
\end{equation}
where $r$ is the distance between the reference point and the center of the exchange hole. $\zeta$ is a positive real number and represents the width of the hole. In the original BR approach\cite{rf:becke1989pra} the factor $A$ in Eq. (\ref{eq:GBR}) is determined as a function of $\zeta$ to ensure that the hole population is normalized to one. In contrast, in the GBR method the factor $A$ is left as a flexible parameter which can accommodate the fragmentation of the exchange hole associated with a bond dissociation. Explicitly, $A$ is determined so that the Coulomb potential of the hole given by Eq. (\ref{eq:GBR}) realizes the exact exchange energy density $U_{x}^{\text{exact}}(\bm{r})$ of the real system
\begin{equation}
U_{x}^{\text{exact}}(\bm{r})=\frac{1}{2} \int d \bm{r}^{\prime} \frac{1}{\left|\bm{r}-\bm{r}^{\prime}\right|} n_{x}\left(\bm{r}, \bm{r}^{\prime}\right)
\label{eq:Ux_exact}    
\end{equation}
where $n_{x}\left(\bm{r}, \bm{r}^{\prime}\right)$ stands for the exchange hole associated with $\bm{r}$. In addition, we have two additional conditions to obtain the other parameters $(\zeta, r)$, for which we refer the readers to Ref. \onlinecite{rf:becke2003jcp}. It is known that the GBR parameters $(A, \zeta, r)$ can be uniquely determined for any given reference point through a certain numerical approach. It is, then, possible to define the projected hole localized at a reference point $\bm{r}$. The population $P^\text{GBR}(\bm{r})$ of the projected hole is evaluated as $8\pi A/\zeta^3$. Obviously, $P^\text{GBR}(\bm{r})$ is relevant to the degree of the fragmentation of the real exchange hole. Actually, for an H$_2$ molecule, it is anticipated that $P^\text{GBR}(\bm{r})$ becomes $0.5$ when $R \to \infty$. Hence, a natural choice of the delocalization factor $\mathcal{D}(\bm{r})$ in terms of the GBR exchange hole model is to take
\begin{equation}
\mathcal{D}(\bm{r}) = N_s(1-P^\text{GBR}(\bm{r}))
\label{eq:dlcf1}
\end{equation} 
where $N_s$ is the number of sites onto which the exchange hole is to be fragmented and $N_s = 2$ is supposed for the dissociation limit of a diatomic molecule. In more general approach that can be applied to the multi-site fragmentation it would be appropriate to take $N_{s}$ as the inverse of weighted average of $P^\text{GBR}(\bm{r})$, thus,  
\begin{equation}
N_s = \left( \frac{1}{N} \int d\bm{r} P^\text{GBR}(\bm{r})n(\bm{r}) \right)^{-1} 
\label{eq:dlcf2}
\end{equation} 
where $N$ is the number of electrons with the spin $\sigma$. The delocalization factor $\mathcal{D}^e(\epsilon)$ dependent on the energy coordinate $\epsilon$ is merely obtained by the weighted average over the isosurface with coordinate $\epsilon$, 
\begin{equation}
\mathcal{D}^{e}(\epsilon)=n^e(\epsilon)^{-1} \int d \bm{r} \mathcal{D}(\bm{r}) n(\bm{r}) \delta(\epsilon-\upsilon(\bm{r}))
\label{eq:D_e}
\end{equation}
which can be justified when $\mathcal{D}(\bm{r})$ is almost constant over the isosurfaces of the external potential $\upsilon$.  In Fig. \ref{Deloc} $\mathcal{D}(\bm{r})$ is mapped with colors on the isosurfaces of the potential $\upsilon$ for the various bond distances $R$ of H$_2$. 
It is shown in the figure that $\mathcal{D}(\bm{r})$ increases with increasing bond distance $R$ as expected. At each bond distance it is also found that $\mathcal{D}(\bm{r})$ tends to have relatively larger values near the nuclei, that is, $\mathcal{D}(\bm{r})$ is large in the region where the coordinate $\epsilon$ has larger values. It is also interesting to note that $\mathcal{D}(\bm{r})$ is relatively smaller in the spatial region between the nuclei. The projection of Eq. (\ref{eq:D_e}) can be justified when $\mathcal{D}(\bm{r})$ is almost constant over the isosurfaces of the external potential $\upsilon$. When $R$ is large, it seems that $\mathcal{D}(\bm{r})$ is nearly constant on an isosurface. On the other hand, for smaller $R$ the constancy of $\mathcal{D}(\bm{r})$ is not fulfilled on the isosurfaces with small $\upsilon$ in particular. However, $\mathcal{D}(\bm{r})$ itself stays small in the region where both $R$ and $\upsilon$ are small. Futhermore, the energy electron density $n^e(\epsilon)$ is also small on the smaller coordinate $\epsilon$. Therefore, the error due to Eq. (\ref{eq:D_e}) is supposed to be small by consulting Eq. (\ref{eq:Exc_static_rev}). In Fig. \ref{Deloc} number of sites $N_s$ given by Eq. (\ref{eq:dlcf2}) are also presented, which shows $N_s$ increases with increasing distance $R$ as expected. In principle, $N_s$ becomes $1.0$ and $2.0$ when $R = 0$ and $\infty$, respectively.  \\
\indent For the correlation function $F(\mathcal{D}^e(\epsilon))$ in Eq. (\ref{eq:Exc_static_rev}), we first tested the simplest model that $F(\mathcal{D}^e(\epsilon)) = \mathcal{D}^e(\epsilon)$. However, we found it underestimates $E_{sc}$ in the intermediate region of $R$ as will be shown later. Thus, we introduce upward-convex polynomial functions $F_k(\mathcal{D}^e(\epsilon))$
\begin{equation}
F_k(\mathcal{D}^e(\epsilon)) = 1-(1-\mathcal{D}^e(\epsilon))^k \;\;\;  (k=0,1, \cdots)
\label{eq:F}
\end{equation}
Specifically, $F_0$ is identically zero and $F_k$ asymptotically goes to 1 when $k \to \infty$.  Thus, the series of the polynomial functions $F_k$ covers the full range of the correlation strength.  It is also important to note that the relations $F_k(0)=0$ and $F_k(1)=1$ is guaranteed for any choice of $k \ge 1$. In this work, the order $k$ solely serves as an adjustable parameter in the numerical implementation of Eq. (\ref{eq:Exc_static_rev}) besides those intrinsic to the BLYP functional. Although $k$ can be a real number, we only consider a positive integer because a precise optimization of the functional is not the major objective of this work. $0 \le \mathcal{D}^e(\epsilon) \le 1$ is supposed in principle for diatomic system. However, in the case of the fragmentation of the hole into multiple sites $\mathcal{D}^e(\epsilon)$ possibly becomes larger than 1. Hence, $F_k (x) = 1.0\; (1 < x)$ has to be imposed on $F_k (x) (k = 1, 2, \cdots) $. \\
\indent For the numerical calculations we utilized \textquoteleft Vmol\textquoteright\; program\cite{rf:takahashi2000cl, takahashi2001jpca, rf:takahashi2001jcc} revised for the present development. The program is based on the real-space grid approach\cite{rf:chelikowsky1994prb, rf:chelikowsky1994prl} where the interaction between valence electrons and nuclei is represented with non-local pseudopotentials in the Kleinman and Bylander separable form\cite{rf:kleinman1982prl}. The charges of the core electrons are summed to the nuclear charge to evaluate the energy coordinate for the valence electrons.  The energy distribution functions $n^e(\epsilon)$ is constructed for the densities on the discrete grid points. To increase the sampling points we introduced a double grid technique\cite{rf:ono1999prl} where the density and its derivatives on a dense grid are obtained through the 4th-order polynomial interpolations of the values on the coarse grids.  The number of the double grids between the coarse grids is set at $N_\text{den}=5$. As was done in the previous paper\cite{rf:Takahashi2018}, the axis for the energy coordinate $\epsilon$ was uniformly descretized with $N^e_1$ points in the lower energy region($\epsilon_\text{min} \le \epsilon \le \epsilon_\text{core}$), while $N^e_2$ points were uniformly placed in the log-scaled coordinate in the higher energy region($\epsilon_\text{core} \le \epsilon \le \epsilon_\text{max}$). The explicit values for $(N_1^e, N_2^e)$ and $(\epsilon_\text{min}, \epsilon_\text{core}, \epsilon_\text{max})$ as well as the method to discretize the energy coordinate are presented in \textquoteleft Supplemental Material\textquoteright. The reference electron density $n_0$ was build with the LDA functional and the BLYP functional was utilized in the functional of Eq. (\ref{eq:Exc_static_rev}).    
\section{\label{sec:level4}Results and Discussion}
\subsection{Bond Dissociations}
\begin{figure}[h]
\centering
\scalebox{0.24}[0.24] {\includegraphics[trim=10 10 30 10,clip]{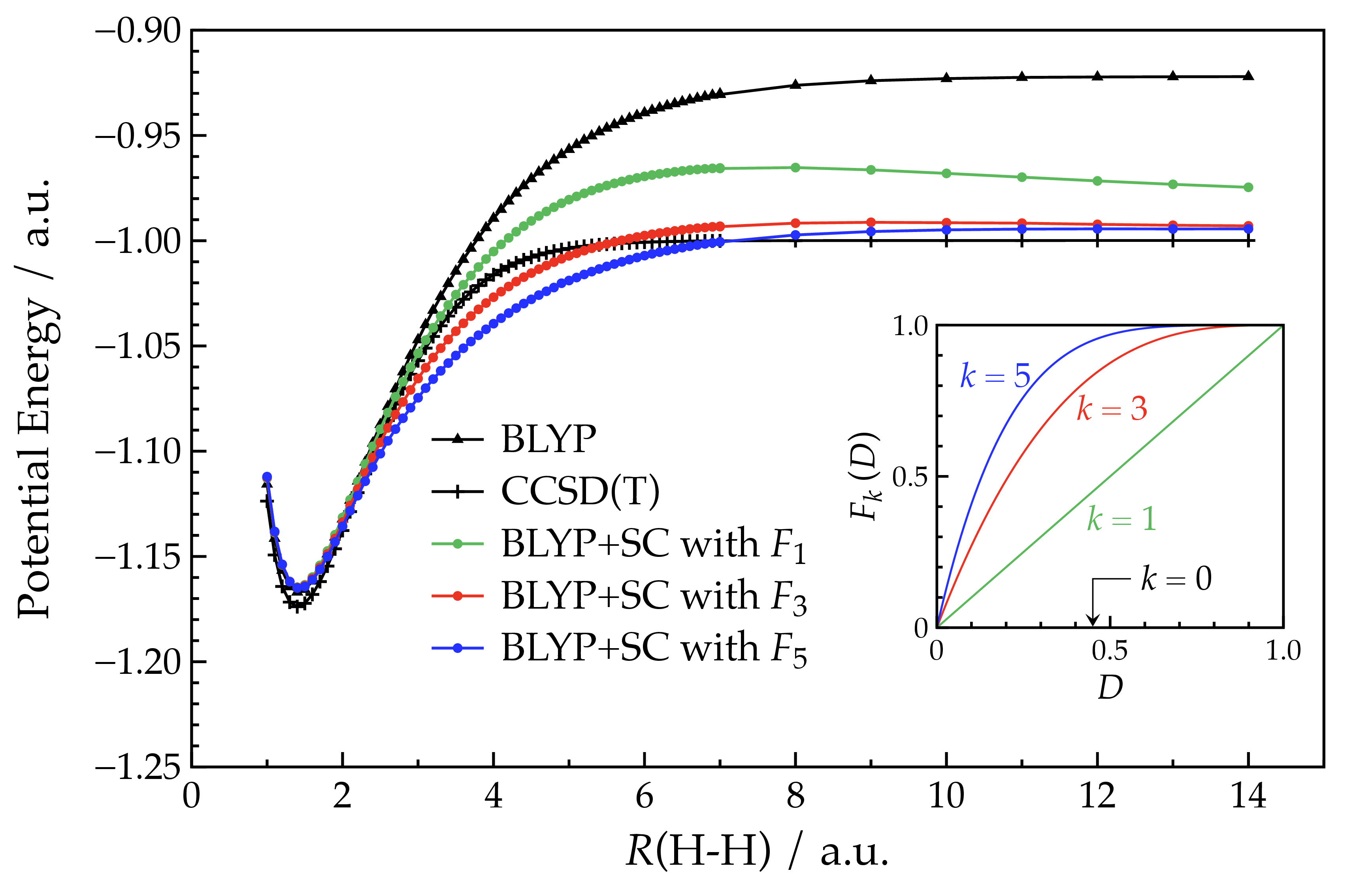}}            
\caption{\label{h2} Potential energy curves for dissociating H$_2$ molecule. The graphs for the correlation factors $F_k$ of Eq. (\ref{eq:F}) as functions of the delocalization factor $\mathcal{D}^e(\epsilon)$(Eq. (\ref{eq:D_e})) are superposed. }
\end{figure}

\begin{figure}[h!]
\centering
\scalebox{0.24}[0.24] {\includegraphics[trim=10 10 20 10,clip]{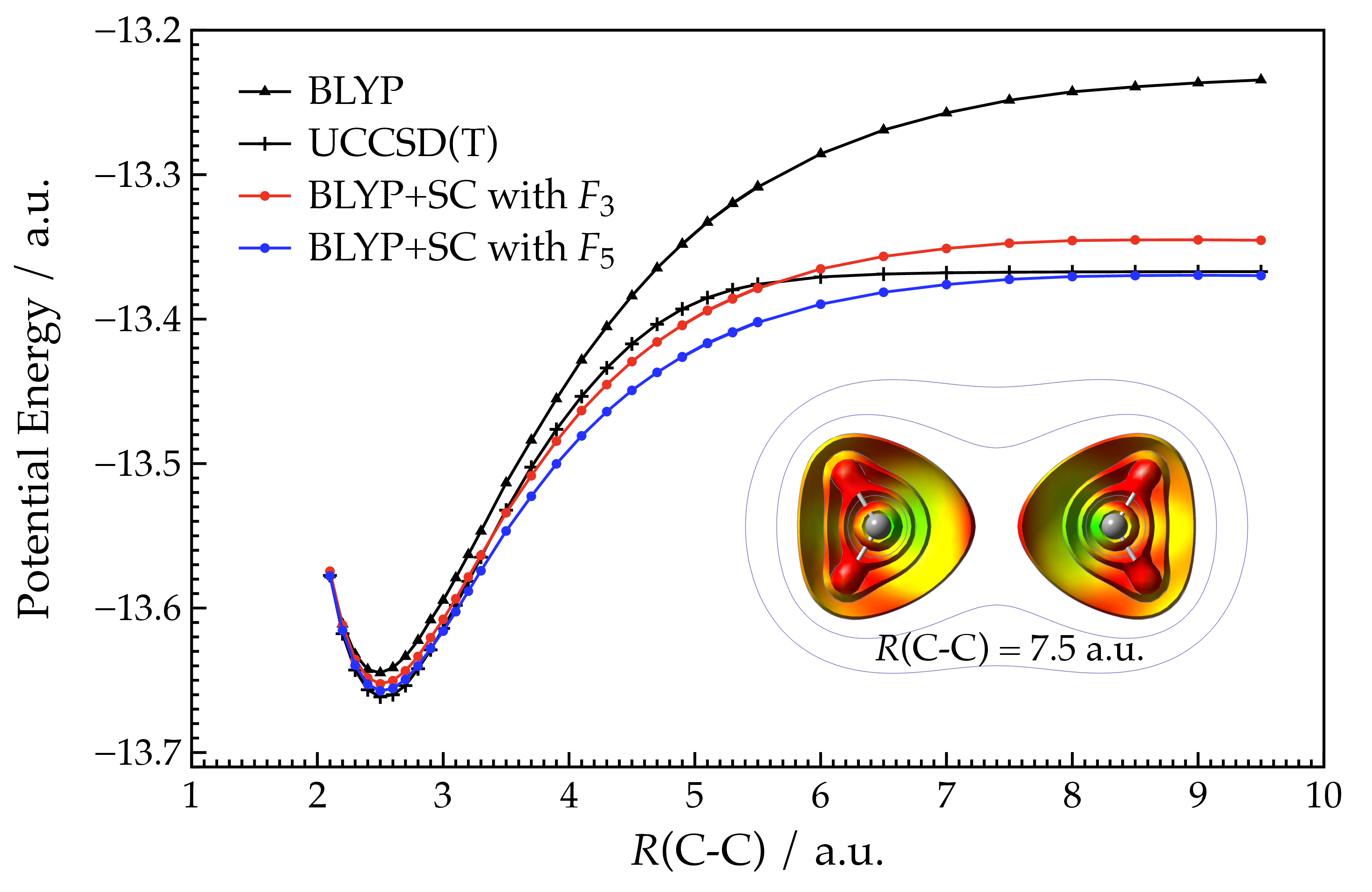}}            
\caption{\label{c2h4} Potential energy curves for the dissociating double bond in C$_2$H$_4$. The delocalization factor $\mathcal{D}(\bm{r})$ is depicted in the inset. The color assignment for $\mathcal{D}(\bm{r})$ is common to that for Fig. \ref{Deloc}. The contour surfaces of $\upsilon(\bm{r})$ are for 3.0, 4.0, 5.0, and 7.0 a.u. and the contour lines are for 2.0, 2.5 a.u. All the potential energy curves are shifted so that the energies of the left ends of the graphs match that of \textquoteleft BLYP+SC with $F_5$\textquoteright.  }
\end{figure}

\begin{figure}[h]
\centering
\scalebox{0.24}[0.24] {\includegraphics[trim=10 10 20 10,clip]{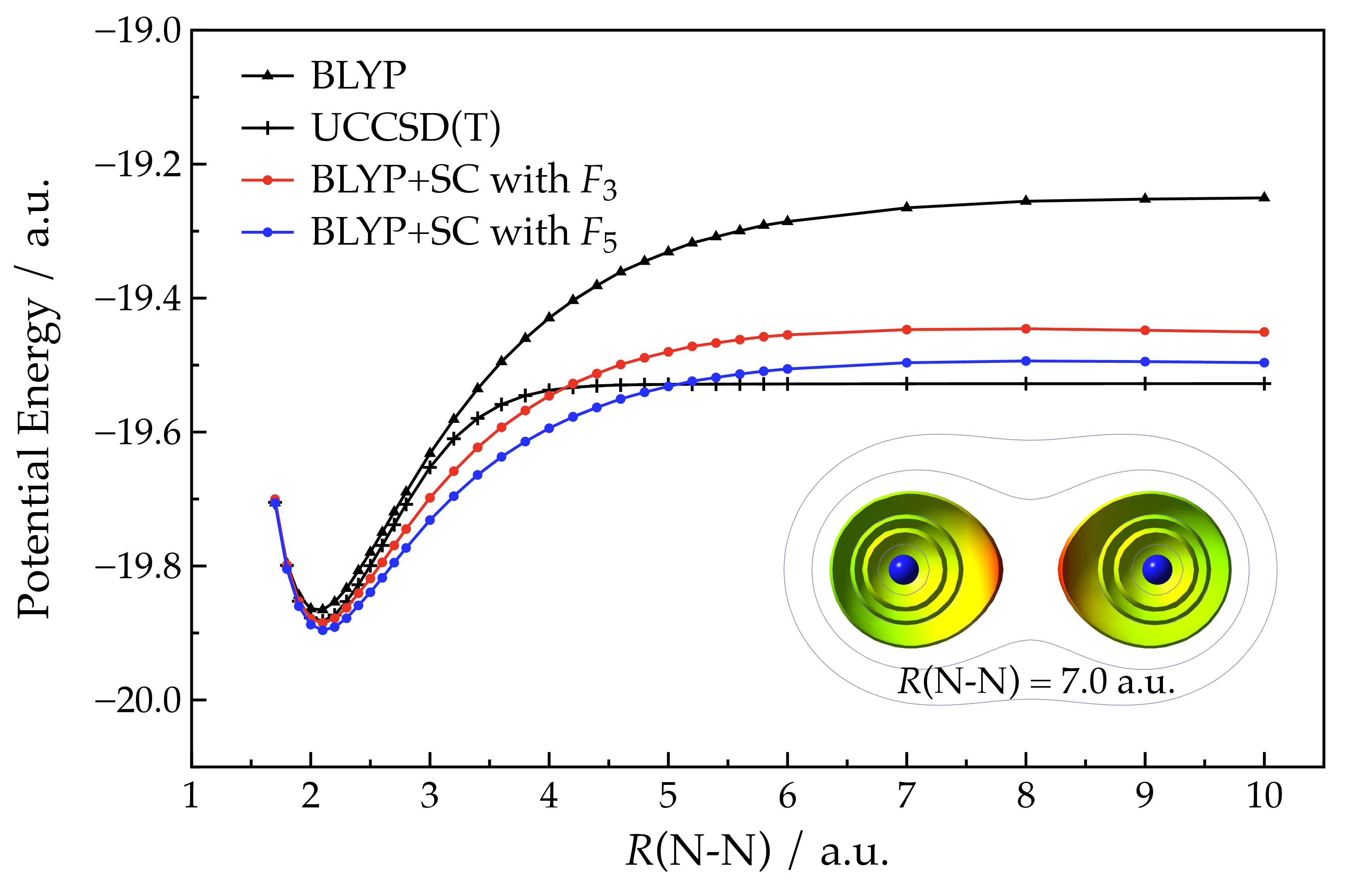}}            
\caption{\label{n2} Potential energy curves for the dissociating triple bond in N$_2$. The delocalization factor $\mathcal{D}(\bm{r})$ is depicted in the inset. The color assignment for $\mathcal{D}(\bm{r})$ is common to that for Fig. \ref{Deloc}. The contour surfaces of $\upsilon(\bm{r})$ are for 3.0, 4.0, and 5.0 a.u. and the contour lines are for 2.0, 2.5, and 8.0 a.u. All the potential energy curves are shifted so that the energies of the left ends of the graphs match that of \textquoteleft BLYP+SC with $F_5$\textquoteright.  }
\end{figure}

We first applied Eq. (\ref{eq:Exc_static_rev}) to the dissociation of an H$_2$ molecule. Here, we leave the sole adjustable parameter $k$ in the correlation factor(Eq. (\ref{eq:F})) undetermined. We performed a series of calculations with various $k$ values including pure BLYP functional based on the conventional DFT. As a reference we also carried out CCSD(T) calculation by conducting \textquoteleft Gaussian 09\textquoteright\; program package, where aug-cc-pVQZ basis set was employed. The results are shown in Fig. \ref{h2} where the graphs for $F_k$ are superposed. It is shown that the BLYP energy with the static correlation (SC) using factor $F_1$ does not fully reproduce the dissociation limit though it provides a significant improvement on the pure BLYP functional. The larger correlation factors $F_3$ and $F_5$ almost successfully realize the dissociation behavior though factor $F_5$ rather overestimates SC in the intermediate H$-$H distance. Actually, dissociation energies $E_\text{d}$ are, respectively, computed as 107.8 and 107.1 kcal/mol with these factors, showing good agreements with the CCSD(T) calculation (109.2 kcal/mol). Thus, we determined the parameter as $k=3$ in the following applications.   \\
\indent Next, we examined the dissociation of a double bond in C$_2$H$_4$. To this end the bond length is gradually increased with the geometry of the CH$_2$ groups being fixed at that of the equilibrium structure in ethene. In stretching the bond we notice that an anti-bonding $\sigma^*$  orbital comes in the set of the low-lying $N_\text{val}$ orbitals in place of the $\pi$ orbital, which causes numerical instabilities in the potential energy(PE). To avoid such an unfavorable situation the wave function obtained for a C$-$C distance is used as the initial guess for the SCF of the next ethene configuration created by a small increment of the bond length. The results are presented in Fig. \ref{c2h4} where the delocalization factor $\mathcal{D}(\bm{r})$ mapped onto the isosurfaces of $\upsilon(\bm{r})$ are also shown in the inset. It is noticeable that the use of the correlation factor $F_5$ gives the dissociation energy of $180.4$ kcal/mol showing a good agreement with that given by UCCSD(T)/cc-pVQZ calculation($184.7$ kcal/mol). It is found, however, that the use of the factor $F_3$ fails to attain the dissociation limit of $F_5$ although the behavior of PE in the intermediate region is more favorable. This is simply because $\mathcal{D}(\bm{r})$ is evaluated to be rather small even at $R$(C$-$C) = 7.5 a.u. as shown in the inset. The population of the projected hole by GBR approach is anticipated to be 0.5 on each site at the dissociation limit. Such a situation is almost realized in H$_2$ as shown in Fig. \ref{Deloc}. However, in other systems the hole does not perform ideally as suggested in Refs. \onlinecite{rf:becke2013jcp, rf:kong2016jctc}. That is, the hole population tends to be overestimated, which gives rise to the underestimation of $\mathcal{D}(\bm{r})$ according to Eq. (\ref{eq:dlcf1}). As a consequence the number of site $N_s$ of Eq. (\ref{eq:dlcf2}) will also be underestimated. Actually, $N_s$ is estimated to be only $1.124$ at $R$(C$-$C) = 9.5 a.u. Thus, the major role of the correlation factor $F_k$ is to compensate the drawback. In other words, another method to determine $\mathcal{D}(\bm{r})$ will possibly improve the functional. \\  
\indent Another severe test was performed by studying the dissociation of the triple bond in an N$_2$ molecule. The results are shown in Fig. \ref{n2}. As a reference the result given by UCCSD(T)/cc-pVQZ calculation is also presented. From the UCCSD(T) potential energy curve, dissociation energy $E_\text{d}$ is estimated to be 222.4 kcal/mol. 
In contrast, the present calculations with factors $F_3$ and $F_5$ provide $E_\text{d} = 272.5$ and $250.7$ kcal/mol, respectively. Thus, it is revealed that the present approach overestimates $E_\text{d}$ of the triple bond. It is also notable that our method underestimates PE of the intermediate N$-$N distances. As shown in the inset the values of $\mathcal{D}(\bm{r})$ are again estimated to be rather small although they are larger than those obtained for C$_2$H$_4$ at $R$(C$-$C) = 7.5 a.u. This may cause the excess rise in the PE of the dissociating region. However, it is quite curious that even the factor $F_5$ could not fully compensate the error as it performed nicely for C$_2$H$_4$. Anyway, it was demonstrated that the present approach with $F_5$ reduces $E_\text{d}$ of the spin-restricted BLYP by $\sim 136$ kcal/mol, showing a better performance than B13\cite{rf:becke2013jcp} and KP\cite{rf:kong2016jctc} functionals.       
 \\
\indent As a overall remark the functional of Eq. (\ref{eq:Exc_static_rev}) is almost successful in realizing the correct $E_\text{d}$ of the various chemical bonds. We found, however, the static correlations for the intermediately dissociated bonds tend to be overestimated, which leads artificial lowering of the PE. The error might be attributed to the underestimation of $\mathcal{D}(\bm{r})$ even in the molecules with sufficiently stretched bonds. Thus, a strong correlation factor $F(\epsilon)$ in Eq. (\ref{eq:Exc_static_rev}) is needed to compensate the drawback, which unexpectedly lowers the exchange energies in the intermediate bond distances. Sophistication of the method to evaluate $\mathcal{D}(\bm{r})$ would improve the overall behavior of PE.  
\subsection{Spin-depolarization and Symmetrization}
\begin{figure}[h]
\centering
\scalebox{0.255}[0.255] {\includegraphics[trim=10 10 20 10,clip]{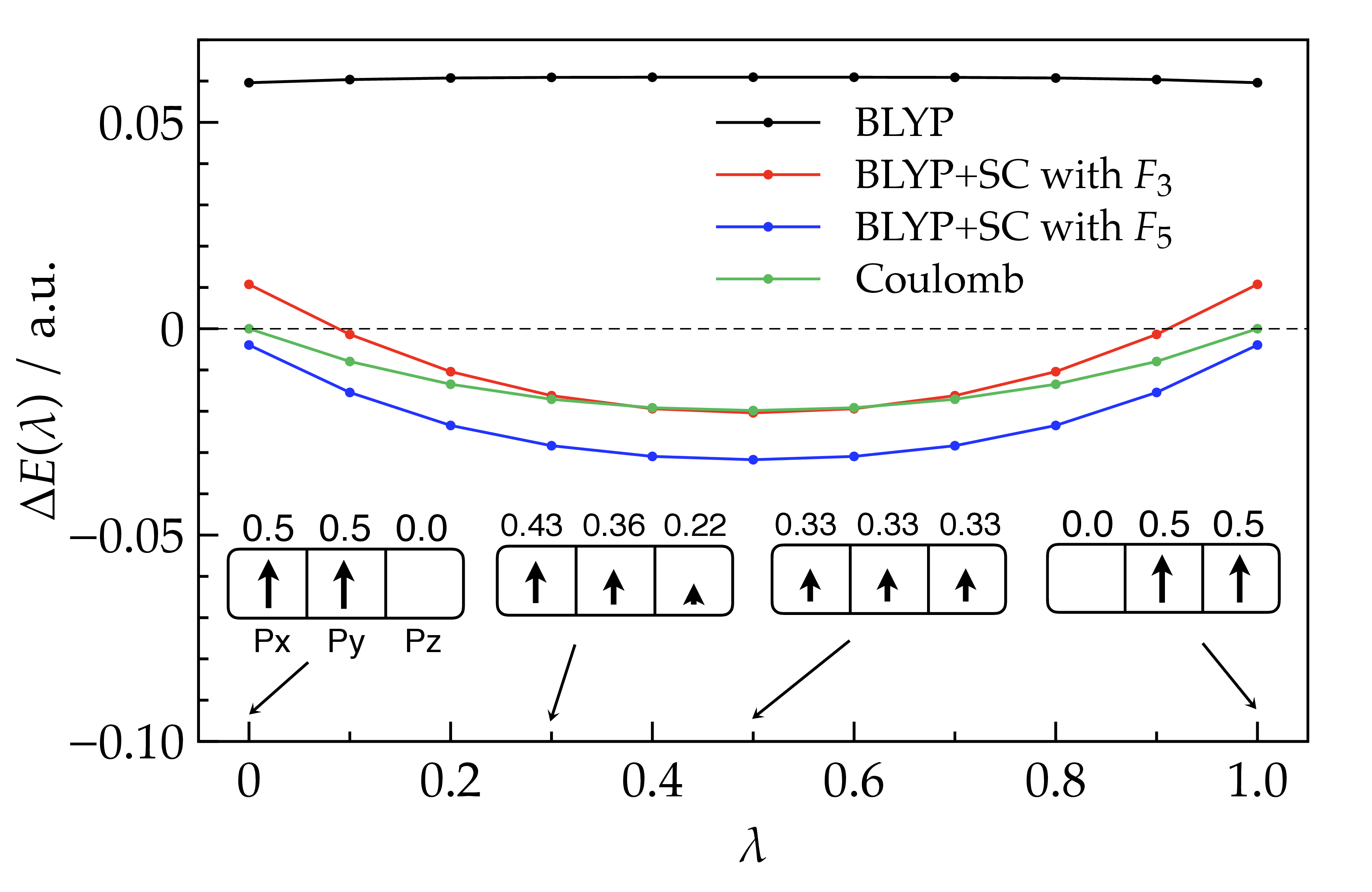}}            
\caption{\label{C_atom} Energy differences $\Delta E$ between total energies of the spin restricted and unrestricted densities of Carbon atom are shown as functions of the mixing parameter $\lambda$. The 2p orbitals in Carbon atom are depicted as the set of three boxes, in which $\alpha$ spins in the spin restricted density are represented with upward bold arrows. The value on each box shows the spin population dependent on the mixing parameter. It is supposed that $\beta$ spin (not shown) has the same population as $\alpha$ on each orbital.  }
\end{figure}

It is proved in Ref. \onlinecite{rf:Yang2000prl} that the exact functional $E_\text{exact}$ for the total energy satisfies the following relation,   
\begin{equation}
E_\text{exact}\left[n_{\text{mix}}\right]=E_{0}^{\upsilon}(N)
\label{eq:strict}
\end{equation}
where $E_{0}^{\upsilon}(N)$ is the ground state energy of the $N$ electrons subject to the external potential $\upsilon$. $n_\text{mix}$ is the linear combination of the $g$-degenerate ground state densities $\{n_{i}\}(i=1,2,\cdots, g)$ for $\upsilon$, 
\begin{equation}
n_{\text{mix}}=\sum_{i=1}^{g} C_{i} n_{i}
\end{equation}
with the normalization condition $\sum_i^{g} C_i = 1$. It should be noted that $n_\text{mix}$ is not $\upsilon$-representable in general\cite{rf:Levy1982pra} and assessing the realization of Eq. (\ref{eq:strict}) in an approximate functional is known as a stringent test for strong correlation\cite{rf:becke2013jcp}.  To examine the performance of the functional of Eq. (\ref{eq:Exc_static_rev}) we construct $n_\text{mix}$ by mixing the three degenerate densities($g=3$) of ground state Carbon atom. We suppose that each density $n_i$ is \textit{spin depolarized}, that is, 2p-shell occupancies of $(2p_x\alpha)^{0.5}(2p_y\alpha)^{0.5}(2p_z\alpha)^{0}(2p_x\beta)^{0.5}(2p_y\beta)^{0.5}(2p_z\beta)^{0}$ are assumed in a component density. Then, a mixing parameter $\lambda (0\le \lambda \le 1)$ is introduced so that the three densities give the equivalent contributions at $\lambda = 1/2$. Explicitly, we adopt $C_1 = \lambda^q, C_2 = (1-\lambda)^q, C_3=1-C_1-C_2$ with $q= \frac{\ln3}{\ln2}$. Accordingly, the occupancy of each p orbital is given as $\frac{1}{2}(1-C_i^\lambda)$. It is readily recognized that in the densities $n_\text{mix}^{\lambda = 0, 1}$ two of the three p orbitals are individually occupied with an electron (0.5 electron with spin $\alpha$ and 0.5 electron with spin $\beta$ on each orbital), while the three orbitals have equivalent occupancies at $\lambda = 1/2$. In Fig. \ref{C_atom} we provide the energy difference $\Delta E(\lambda)$ between the total energy of  Carbon atom with the spin polarized density, namely $(2p_x\alpha)^{1}(2p_y\alpha)^{1}(2p_z\alpha)^{0}(2p_x\beta)^{0}(2p_y\beta)^{0}(2p_z\beta)^{0}$ and the energy $E[n_\text{mix}^\lambda]$ computed with Eq. (\ref{eq:Exc_static_rev}). The exact functional realizes $\Delta E(\lambda) = 0$ irrespective of $\lambda$. In applying Eq. (\ref{eq:Exc_static_rev}) we adopt the energy density $n^e$ of the supermolecule as the reference, which consists of infinitely separated Carbon atoms with the spin polarized densities. The result by the conventional BLYP functional is also presented to make comparisons. The Coulomb interaction $J[n_\text{mix}^\lambda] = \frac{1}{2} \int d\bm{r}d\bm{r}^\prime  \frac{n_\text{mix}^\lambda(\bm{r})n_\text{mix}^\lambda(\bm{r}^\prime)}{|\bm{r}-\bm{r}^\prime|}$ is also shown where $J[n_\text{mix}^{\lambda=0}]=J[n_\text{mix}^{\lambda=1}]$ is taken as the standard.   \\
\indent The specific energy differences $\Delta E(0)$ or $\Delta E(1)$ show the energy $\Delta E_\text{depol}$ due to the spin depolarization, while $\Delta E(1/2)$ subtracted by $\Delta E_\text{depol}$ is the energy $\Delta E_\text{sym}$ due to symmetrization, i.e. $\Delta E_\text{sym} = \Delta E(1/2) - \Delta E_\text{depol}$. It is notable in the figure that $\Delta E_\text{depol}$ is estimated to be almost 0 by BLYP+SC(Eq. (\ref{eq:Exc_static_rev})) with the correlation factor $F_5$ although use of the factor $F_3$ slightly degrades the energy. It was demonstrated that the functional works nicely in computing the energy change $\Delta E_\text{depol}$ associated with the spin-depolarization. On the other hand, $\Delta E_\text{depol}$ given by the conventional GGA functional(BLYP) becomes positively much larger ($\sim$ 0.06 a.u.) due to the lack of the static correlation. As shown in Fig. \ref{C_atom} the Coulomb interaction $J[n_\text{mix}^{\lambda}]$ decreases with the increase in $\lambda$ from $0$ to $1/2$ since the electron density $n_\text{mix}^{\lambda}$ fully delocalizes at $\lambda = 1/2$ due to the perfect symmetrization in space. $\Delta E(\lambda)$ of Eq. (\ref{eq:Exc_static_rev}) with the factor $F_5$ unfavorably decreases along with $J[n_\text{mix}^{\lambda}]$ suggesting $E_{xc}^e[n_\text{mix}^{\lambda}]$ itself stays almost constant. The energy $\Delta E(\lambda)$ given by the factor $F_3$ shows almost the same tendency as $F_5$. Although the values $\Delta E(1/2)$ by $F_3$ as well as $F_5$ are much closer to 0 than that computed with the BLYP functional, $\Delta E_\text{sym}$ itself by Eq. (\ref{eq:Exc_static_rev}) is larger than that with the BLYP functional. The constancy of $\Delta E(\lambda)$ with BLYP implies that the decrease in $J[n_\text{mix}^{\lambda}]$ is perfectly compensated with the increase in the $E_{xc}[n_\text{mix}^{\lambda}]$. The problem in $\Delta E_\text{sym}$ has a relevance with the other error in DFT referred to as self-interaction error(SIE) and should be considered in the forthcoming issues.  
 \\
\section{\label{sec:level5}Conclusion}
We developed an electron density functional to compute strong electron correlation referred to as static correlation(SC) that emerges when a covalent bond is split. In contrast to the conventional DFT the energy electron distribution $n^e(\epsilon)$ instead of $n(\bm{r})$ plays a fundamental role in the functional development on the basis of the rigorous framework of DFT as discussed in Ref. \onlinecite{rf:Takahashi2018}. The key feature of the new distribution is that $n^e_\lambda(\epsilon)$ stays constant with respect to the variation of parameter $\lambda$ which  couples the spin polarized density with that for the entangled state when the fragments are infinitely separated in a homolytic dissociation. \\
\indent A remarkable progress was made in the functional by introducing the correlation factor as a function of the degree $\mathcal{D}^e(\epsilon)$ of delocalization of the exchange hole on the coordinate $\epsilon$. In the present work, $\mathcal{D}^e(\epsilon)$ is obtained by utilizing the generalized Becke-Roussel(GBR) machinery.   \\
\indent The functional was applied to the computation of the splits of single, double and triple bonds. Namely, the dissociations of the bonds in H$_2$, C$_2$H$_4$, and N$_2$ were studied. It was demonstrated that the dissociations of the single and the double bonds are faithfully realized with the correlation factor $F_5$ showing reasonable agreements with those obtained by sophisticated molecular orbital theories. However, the dissociation energy was found to be overestimated in N$_2$ although substantially improved as compared with the conventional GGA functional. The source of the discrepancy might be attributed to the underestimation in $\mathcal{D}^e(\epsilon)$ by means of the GBR method. The present functional was also applied to the computation of the energy change due to the spin depolarization and symmetrization in Carbon atom, which showed a remarkable improvement over the conventional one.   \\
\indent The present work suggests the possibility of a paradigm for the electronic density functional based on the energy coordinate $\epsilon$. In the conventional DFT functional the electronic correlation of an electron at $\bm{r}$ is solely determined by the \textit{local} properties, namely, the electron density and its derivatives at $\bm{r}$. This condition will impose a severe limitation on the applicability of functional. The present DFT approach in terms of the energy electron density offers a simple framework to overcome the problem due to the spatial locality intrinsic to the LDA-based DFT.  The development of the kinetic energy functional is undoubtedly the toughest challenge in DFT, where inclusion of a non-local term is considered to be essential as discussed in Ref. \onlinecite{rf:Wang1992prb}. The use of the energy coordinate instead of $\bm{r}$ in DFT will serve a potential approach to a kinetic energy functional.         
\begin{acknowledgments}
This paper was supported by the Grant-in-Aid for Scientific Research on Innovative Areas (No. 23118701) from the Ministry of Education, Culture, Sports, Science, and Technology (MEXT); the Grant-in-Aid for Challenging Exploratory Research (No. 25620004) from the Japan Society for the Promotion of Science (JSPS); and the Grant-in-Aid for Scientific Research(C) (No. 17K05138) from the Japan Society for the Promotion of Science (JSPS). This research also used computational resources of the HPCI system provided by Kyoto, Nagoya, and Osaka university through the HPCI System Research Project (Project IDs: hp170046, hp180030, hp180032, hp190011, and hp200016).
\end{acknowledgments}  

\bibliography{apssamp}

\providecommand{\noopsort}[1]{}\providecommand{\singleletter}[1]{#1}%
\begin{thebibliography}{21}%
\makeatletter
\providecommand \@ifxundefined [1]{%
 \@ifx{#1\undefined}
}%
\providecommand \@ifnum [1]{%
 \ifnum #1\expandafter \@firstoftwo
 \else \expandafter \@secondoftwo
 \fi
}%
\providecommand \@ifx [1]{%
 \ifx #1\expandafter \@firstoftwo
 \else \expandafter \@secondoftwo
 \fi
}%
\providecommand \natexlab [1]{#1}%
\providecommand \enquote  [1]{``#1''}%
\providecommand \bibnamefont  [1]{#1}%
\providecommand \bibfnamefont [1]{#1}%
\providecommand \citenamefont [1]{#1}%
\providecommand \href@noop [0]{\@secondoftwo}%
\providecommand \href [0]{\begingroup \@sanitize@url \@href}%
\providecommand \@href[1]{\@@startlink{#1}\@@href}%
\providecommand \@@href[1]{\endgroup#1\@@endlink}%
\providecommand \@sanitize@url [0]{\catcode `\\12\catcode `\$12\catcode
  `\&12\catcode `\#12\catcode `\^12\catcode `\_12\catcode `\%12\relax}%
\providecommand \@@startlink[1]{}%
\providecommand \@@endlink[0]{}%
\providecommand \url  [0]{\begingroup\@sanitize@url \@url }%
\providecommand \@url [1]{\endgroup\@href {#1}{\urlprefix }}%
\providecommand \urlprefix  [0]{URL }%
\providecommand \Eprint [0]{\href }%
\providecommand \doibase [0]{https://doi.org/}%
\providecommand \selectlanguage [0]{\@gobble}%
\providecommand \bibinfo  [0]{\@secondoftwo}%
\providecommand \bibfield  [0]{\@secondoftwo}%
\providecommand \translation [1]{[#1]}%
\providecommand \BibitemOpen [0]{}%
\providecommand \bibitemStop [0]{}%
\providecommand \bibitemNoStop [0]{.\EOS\space}%
\providecommand \EOS [0]{\spacefactor3000\relax}%
\providecommand \BibitemShut  [1]{\csname bibitem#1\endcsname}%
\let\auto@bib@innerbib\@empty
\bibitem [{\citenamefont {Hansen}\ and\ \citenamefont
  {Mcdonald}(2013)}]{rf:Hansen2013}%
  \BibitemOpen
  \bibfield  {author} {\bibinfo {author} {\bibfnamefont {J.-P.}\ \bibnamefont
  {Hansen}}\ and\ \bibinfo {author} {\bibfnamefont {I.~R.}\ \bibnamefont
  {Mcdonald}},\ }\href@noop {} {\emph {\bibinfo {title} {Theory of Simple
  Liquids}}}\ (\bibinfo  {publisher} {Academic Press},\ \bibinfo {year}
  {2013})\BibitemShut {NoStop}%
\bibitem [{\citenamefont {Parr}\ and\ \citenamefont
  {Yang}(1989)}]{rf:parr_yang_eng}%
  \BibitemOpen
  \bibfield  {author} {\bibinfo {author} {\bibfnamefont {R.~G.}\ \bibnamefont
  {Parr}}\ and\ \bibinfo {author} {\bibfnamefont {W.}~\bibnamefont {Yang}},\
  }\href@noop {} {\emph {\bibinfo {title} {Density-functional theory of atoms
  and molecules}}}\ (\bibinfo  {publisher} {Oxford university press},\ \bibinfo
  {address} {New York},\ \bibinfo {year} {1989})\BibitemShut {NoStop}%
\bibitem [{\citenamefont {Kohn}\ and\ \citenamefont
  {Sham}(1965)}]{rf:kohn1965pr}%
  \BibitemOpen
  \bibfield  {author} {\bibinfo {author} {\bibfnamefont {W.}~\bibnamefont
  {Kohn}}\ and\ \bibinfo {author} {\bibfnamefont {L.~J.}\ \bibnamefont
  {Sham}},\ }\bibfield  {title} {\bibinfo {title} {{Self-consistent equations
  including exchange and correlation effects}},\ }\href@noop {} {\bibfield
  {journal} {\bibinfo  {journal} {Phys. Rev.}\ }\textbf {\bibinfo {volume}
  {140}},\ \bibinfo {pages} {A1133} (\bibinfo {year} {1965})}\BibitemShut
  {NoStop}%
\bibitem [{\citenamefont {Cohen}\ \emph {et~al.}(2008)\citenamefont {Cohen},
  \citenamefont {Mori-S{\'{a}}nchez},\ and\ \citenamefont {Yang}}]{Cohen2008}%
  \BibitemOpen
  \bibfield  {author} {\bibinfo {author} {\bibfnamefont {A.~J.}\ \bibnamefont
  {Cohen}}, \bibinfo {author} {\bibfnamefont {P.}~\bibnamefont
  {Mori-S{\'{a}}nchez}},\ and\ \bibinfo {author} {\bibfnamefont
  {W.}~\bibnamefont {Yang}},\ }\bibfield  {title} {\bibinfo {title} {{Insights
  into Current Limitations of Density Functional Theory}},\ }\href
  {https://doi.org/10.1126/science.1158722} {\bibfield  {journal} {\bibinfo
  {journal} {Science}\ }\textbf {\bibinfo {volume} {321}},\ \bibinfo {pages}
  {792} (\bibinfo {year} {2008})}\BibitemShut {NoStop}%
\bibitem [{\citenamefont {Becke}(2003)}]{rf:becke2003jcp}%
  \BibitemOpen
  \bibfield  {author} {\bibinfo {author} {\bibfnamefont {A.~D.}\ \bibnamefont
  {Becke}},\ }\bibfield  {title} {\bibinfo {title} {A real-space model of
  nondynamical correlation},\ }\href@noop {} {\bibfield  {journal} {\bibinfo
  {journal} {J. Chem. Phys.}\ }\textbf {\bibinfo {volume} {119}},\ \bibinfo
  {pages} {2972} (\bibinfo {year} {2003})}\BibitemShut {NoStop}%
\bibitem [{\citenamefont {Takahashi}(2018)}]{rf:Takahashi2018}%
  \BibitemOpen
  \bibfield  {author} {\bibinfo {author} {\bibfnamefont {H.}~\bibnamefont
  {Takahashi}},\ }\bibfield  {title} {\bibinfo {title} {Density-functional
  theory based on the electron distribution on the energy coordinate},\
  }\href@noop {} {\bibfield  {journal} {\bibinfo  {journal} {J. Phys. B:
  Atomic, Molecular and Optical Physics}\ }\textbf {\bibinfo {volume} {51}},\
  \bibinfo {pages} {055102(11pp)} (\bibinfo {year} {2018})}\BibitemShut
  {NoStop}%
\bibitem [{\citenamefont {Becke}(1988)}]{rf:becke1988pra}%
  \BibitemOpen
  \bibfield  {author} {\bibinfo {author} {\bibfnamefont {A.~D.}\ \bibnamefont
  {Becke}},\ }\bibfield  {title} {\bibinfo {title} {{Density-functional
  exchange-energy approximation with correct asymptotic behavior}},\
  }\href@noop {} {\bibfield  {journal} {\bibinfo  {journal} {Phys. Rev. A}\
  }\textbf {\bibinfo {volume} {38}},\ \bibinfo {pages} {3098} (\bibinfo {year}
  {1988})}\BibitemShut {NoStop}%
\bibitem [{\citenamefont {Lee}\ \emph {et~al.}(1988)\citenamefont {Lee},
  \citenamefont {Yang},\ and\ \citenamefont {Parr}}]{rf:lee1988prb}%
  \BibitemOpen
  \bibfield  {author} {\bibinfo {author} {\bibfnamefont {C.}~\bibnamefont
  {Lee}}, \bibinfo {author} {\bibfnamefont {W.}~\bibnamefont {Yang}},\ and\
  \bibinfo {author} {\bibfnamefont {R.~G.}\ \bibnamefont {Parr}},\ }\bibfield
  {title} {\bibinfo {title} {{Development of the Colle-Salvetti
  correlation-energy formula into a functional of the electron density}},\
  }\href@noop {} {\bibfield  {journal} {\bibinfo  {journal} {Phys. Rev. B}\
  }\textbf {\bibinfo {volume} {37}},\ \bibinfo {pages} {785} (\bibinfo {year}
  {1988})}\BibitemShut {NoStop}%
\bibitem [{\citenamefont {Becke}\ and\ \citenamefont
  {Roussel}(1989)}]{rf:becke1989pra}%
  \BibitemOpen
  \bibfield  {author} {\bibinfo {author} {\bibfnamefont {A.~D.}\ \bibnamefont
  {Becke}}\ and\ \bibinfo {author} {\bibfnamefont {M.~R.}\ \bibnamefont
  {Roussel}},\ }\bibfield  {title} {\bibinfo {title} {Exchange holes in
  inhomogeneous system: a coordinate-space model},\ }\href@noop {} {\bibfield
  {journal} {\bibinfo  {journal} {Phys. Rev. A}\ }\textbf {\bibinfo {volume}
  {39}},\ \bibinfo {pages} {3761} (\bibinfo {year} {1989})}\BibitemShut
  {NoStop}%
\bibitem [{\citenamefont {Takahashi}\ \emph {et~al.}(2000)\citenamefont
  {Takahashi}, \citenamefont {Hori}, \citenamefont {Wakabayashi},\ and\
  \citenamefont {Nitta}}]{rf:takahashi2000cl}%
  \BibitemOpen
  \bibfield  {author} {\bibinfo {author} {\bibfnamefont {H.}~\bibnamefont
  {Takahashi}}, \bibinfo {author} {\bibfnamefont {T.}~\bibnamefont {Hori}},
  \bibinfo {author} {\bibfnamefont {T.}~\bibnamefont {Wakabayashi}},\ and\
  \bibinfo {author} {\bibfnamefont {T.}~\bibnamefont {Nitta}},\ }\bibfield
  {title} {\bibinfo {title} {A density functional study for hydrogen bond
  energy by employing real space grids},\ }\href@noop {} {\bibfield  {journal}
  {\bibinfo  {journal} {Chem. Lett.}\ }\textbf {\bibinfo {volume} {3}},\
  \bibinfo {pages} {222} (\bibinfo {year} {2000})}\BibitemShut {NoStop}%
\bibitem [{\citenamefont {Takahashi}\ \emph
  {et~al.}(2001{\natexlab{a}})\citenamefont {Takahashi}, \citenamefont {Hori},
  \citenamefont {Wakabayashi},\ and\ \citenamefont
  {Nitta}}]{takahashi2001jpca}%
  \BibitemOpen
  \bibfield  {author} {\bibinfo {author} {\bibfnamefont {H.}~\bibnamefont
  {Takahashi}}, \bibinfo {author} {\bibfnamefont {T.}~\bibnamefont {Hori}},
  \bibinfo {author} {\bibfnamefont {T.}~\bibnamefont {Wakabayashi}},\ and\
  \bibinfo {author} {\bibfnamefont {T.}~\bibnamefont {Nitta}},\ }\bibfield
  {title} {\bibinfo {title} {{Real Space Ab Initio Molecular Dynamics
  Simulations for the Reactions of OH Radical/OH Anion with Formaldehyde}},\
  }\href@noop {} {\bibfield  {journal} {\bibinfo  {journal} {J. Phys. Chem. A}\
  }\textbf {\bibinfo {volume} {105}},\ \bibinfo {pages} {4351} (\bibinfo {year}
  {2001}{\natexlab{a}})}\BibitemShut {NoStop}%
\bibitem [{\citenamefont {Takahashi}\ \emph
  {et~al.}(2001{\natexlab{b}})\citenamefont {Takahashi}, \citenamefont {Hori},
  \citenamefont {Hashimoto},\ and\ \citenamefont
  {Nitta}}]{rf:takahashi2001jcc}%
  \BibitemOpen
  \bibfield  {author} {\bibinfo {author} {\bibfnamefont {H.}~\bibnamefont
  {Takahashi}}, \bibinfo {author} {\bibfnamefont {T.}~\bibnamefont {Hori}},
  \bibinfo {author} {\bibfnamefont {H.}~\bibnamefont {Hashimoto}},\ and\
  \bibinfo {author} {\bibfnamefont {T.}~\bibnamefont {Nitta}},\ }\bibfield
  {title} {\bibinfo {title} {A hybrid qm/mm method employing real space grids
  for qm water in the tip4p water solvents},\ }\href@noop {} {\bibfield
  {journal} {\bibinfo  {journal} {J. Comp. Chem.}\ }\textbf {\bibinfo {volume}
  {22}},\ \bibinfo {pages} {1252} (\bibinfo {year}
  {2001}{\natexlab{b}})}\BibitemShut {NoStop}%
\bibitem [{\citenamefont {Chelikowsky}\ \emph
  {et~al.}(1994{\natexlab{a}})\citenamefont {Chelikowsky}, \citenamefont
  {Troullier}, \citenamefont {Wu},\ and\ \citenamefont
  {Saad}}]{rf:chelikowsky1994prb}%
  \BibitemOpen
  \bibfield  {author} {\bibinfo {author} {\bibfnamefont {J.~R.}\ \bibnamefont
  {Chelikowsky}}, \bibinfo {author} {\bibfnamefont {N.}~\bibnamefont
  {Troullier}}, \bibinfo {author} {\bibfnamefont {K.}~\bibnamefont {Wu}},\ and\
  \bibinfo {author} {\bibfnamefont {Y.}~\bibnamefont {Saad}},\ }\bibfield
  {title} {\bibinfo {title} {High-order finite-difference pseudopotential
  method: an application to diatomic molecules},\ }\href@noop {} {\bibfield
  {journal} {\bibinfo  {journal} {Phys. Rev. B}\ }\textbf {\bibinfo {volume}
  {50}},\ \bibinfo {pages} {11355} (\bibinfo {year}
  {1994}{\natexlab{a}})}\BibitemShut {NoStop}%
\bibitem [{\citenamefont {Chelikowsky}\ \emph
  {et~al.}(1994{\natexlab{b}})\citenamefont {Chelikowsky}, \citenamefont
  {Troullier},\ and\ \citenamefont {Saad}}]{rf:chelikowsky1994prl}%
  \BibitemOpen
  \bibfield  {author} {\bibinfo {author} {\bibfnamefont {J.~R.}\ \bibnamefont
  {Chelikowsky}}, \bibinfo {author} {\bibfnamefont {N.}~\bibnamefont
  {Troullier}},\ and\ \bibinfo {author} {\bibfnamefont {Y.}~\bibnamefont
  {Saad}},\ }\bibfield  {title} {\bibinfo {title} {Finite-difference
  pseudopotential method: electronic structure calculations without a basis},\
  }\href@noop {} {\bibfield  {journal} {\bibinfo  {journal} {Phys. Rev. Lett.}\
  }\textbf {\bibinfo {volume} {72}},\ \bibinfo {pages} {1240} (\bibinfo {year}
  {1994}{\natexlab{b}})}\BibitemShut {NoStop}%
\bibitem [{\citenamefont {Kleinman}\ and\ \citenamefont
  {Bylander}(1982)}]{rf:kleinman1982prl}%
  \BibitemOpen
  \bibfield  {author} {\bibinfo {author} {\bibfnamefont {L.}~\bibnamefont
  {Kleinman}}\ and\ \bibinfo {author} {\bibfnamefont {D.~M.}\ \bibnamefont
  {Bylander}},\ }\bibfield  {title} {\bibinfo {title} {Efficacious form for
  model pseudopotentials},\ }\href@noop {} {\bibfield  {journal} {\bibinfo
  {journal} {Phys. Rev. Lett.}\ }\textbf {\bibinfo {volume} {48}},\ \bibinfo
  {pages} {1425} (\bibinfo {year} {1982})}\BibitemShut {NoStop}%
\bibitem [{\citenamefont {Ono}\ and\ \citenamefont
  {Hirose}(1999)}]{rf:ono1999prl}%
  \BibitemOpen
  \bibfield  {author} {\bibinfo {author} {\bibfnamefont {T.}~\bibnamefont
  {Ono}}\ and\ \bibinfo {author} {\bibfnamefont {K.}~\bibnamefont {Hirose}},\
  }\bibfield  {title} {\bibinfo {title} {Timesaving double-grid method for
  real-space electronic-structure calculations},\ }\href@noop {} {\bibfield
  {journal} {\bibinfo  {journal} {Phys. Rev. Lett.}\ }\textbf {\bibinfo
  {volume} {82}},\ \bibinfo {pages} {5016} (\bibinfo {year}
  {1999})}\BibitemShut {NoStop}%
\bibitem [{\citenamefont {Becke}(2013)}]{rf:becke2013jcp}%
  \BibitemOpen
  \bibfield  {author} {\bibinfo {author} {\bibfnamefont {A.~D.}\ \bibnamefont
  {Becke}},\ }\bibfield  {title} {\bibinfo {title} {Density functionals for
  static, dynamical, and strong correlation},\ }\href@noop {} {\bibfield
  {journal} {\bibinfo  {journal} {J. Chem. Phys.}\ }\textbf {\bibinfo {volume}
  {138}},\ \bibinfo {pages} {074109(10)} (\bibinfo {year} {2013})}\BibitemShut
  {NoStop}%
\bibitem [{\citenamefont {Kong}\ and\ \citenamefont
  {Proynov}(2016)}]{rf:kong2016jctc}%
  \BibitemOpen
  \bibfield  {author} {\bibinfo {author} {\bibfnamefont {J.}~\bibnamefont
  {Kong}}\ and\ \bibinfo {author} {\bibfnamefont {E.}~\bibnamefont {Proynov}},\
  }\bibfield  {title} {\bibinfo {title} {Density functional model for
  nondynamic and strong correlation},\ }\href@noop {} {\bibfield  {journal}
  {\bibinfo  {journal} {J. Chem. Theory Comput.}\ }\textbf {\bibinfo {volume}
  {12}},\ \bibinfo {pages} {133} (\bibinfo {year} {2016})}\BibitemShut
  {NoStop}%
\bibitem [{\citenamefont {Yang}\ \emph {et~al.}(2000)\citenamefont {Yang},
  \citenamefont {Zhang},\ and\ \citenamefont {Ayers}}]{rf:Yang2000prl}%
  \BibitemOpen
  \bibfield  {author} {\bibinfo {author} {\bibfnamefont {W.}~\bibnamefont
  {Yang}}, \bibinfo {author} {\bibfnamefont {Y.}~\bibnamefont {Zhang}},\ and\
  \bibinfo {author} {\bibfnamefont {P.~W.}\ \bibnamefont {Ayers}},\ }\bibfield
  {title} {\bibinfo {title} {Degenerate ground states and a fractional number
  of electrons in density and reduced density matrix functional theory},\
  }\href@noop {} {\bibfield  {journal} {\bibinfo  {journal} {Phys. Rev. Lett.}\
  }\textbf {\bibinfo {volume} {84}},\ \bibinfo {pages} {5172} (\bibinfo {year}
  {2000})}\BibitemShut {NoStop}%
\bibitem [{\citenamefont {Levy}(1982)}]{rf:Levy1982pra}%
  \BibitemOpen
  \bibfield  {author} {\bibinfo {author} {\bibfnamefont {M.}~\bibnamefont
  {Levy}},\ }\bibfield  {title} {\bibinfo {title} {Electron densities in search
  of hamiltonians},\ }\href@noop {} {\bibfield  {journal} {\bibinfo  {journal}
  {Phys. Rev. A}\ }\textbf {\bibinfo {volume} {26}},\ \bibinfo {pages} {1200}
  (\bibinfo {year} {1982})}\BibitemShut {NoStop}%
\bibitem [{\citenamefont {Wang}\ and\ \citenamefont
  {Teter}(1992)}]{rf:Wang1992prb}%
  \BibitemOpen
  \bibfield  {author} {\bibinfo {author} {\bibfnamefont {L.-W.}\ \bibnamefont
  {Wang}}\ and\ \bibinfo {author} {\bibfnamefont {M.~P.}\ \bibnamefont
  {Teter}},\ }\bibfield  {title} {\bibinfo {title} {Kinetic-energy functional
  of the electron density},\ }\href@noop {} {\bibfield  {journal} {\bibinfo
  {journal} {Phys. Rev. B}\ }\textbf {\bibinfo {volume} {45}},\ \bibinfo
  {pages} {13196} (\bibinfo {year} {1992})}\BibitemShut {NoStop}%
\end{thebibliography}%


\providecommand{\noopsort}[1]{}\providecommand{\singleletter}[1]{#1}%
%

\end{document}